\documentclass[a4paper,11pt]{article}
\pdfoutput=1 

\usepackage{jheppub} 

\usepackage[T1]{fontenc} 

\newcommand{\be}{\begin{equation}}
\newcommand{\ee}{\end{equation}}
\newcommand{\bs}{\begin{subequations}}
\newcommand{\es}{\end{subequations}}
\let\no\nonumber 
\def\noi{\no \\} 
\def\({\left(} 
\def\){\right)} 
\def\a{\alpha}
\def\b{\beta}
\def\k{\kappa}
\def\d{\delta}
\def\D{\Delta}
\def\G{\Gamma}
\def\L{\Lambda}
\def\o{\omega}
\def\O{\Omega}
\def\S{\Sigma}
\def\WW{\mathcal{W}}

\title{Hawking Radiation and Entropic Gravity}

\author{Ian Nagle,\note{For proceedings of the Petersburg Nuclear Physics Institute's 50th Winter School on Theoretical Physics.}}
\affiliation{Petersburg Nuclear Physics Institute,\\Gatchina 188300, St. Petersburg, Russia}

\emailAdd{ianagle@gmail.com}

\abstract{I present a new derivation of Hawking radiation by applying the Crooks fluctuation theorem to the laws of black hole thermodynamics. Then, by analogy with the quantum fluctuation theorem, this allows one to identify microstates contributing to the black hole entropy. These microstates have an evolution operator on the horizon between initial and final states that is related to the entropic gravity proposal. Their full calculation perhaps requires a deeper understanding of nonperturbative Hawking radiation.}

\begin{document} 
\maketitle
\flushbottom

\section{Thermodynamics and Gravity?}
In this short paper I'll begin by briefly reviewing some background material, which can otherwise be a little expansive, in an effort to motivate the subsequent developments. Firstly I feel a need to justify the usefulness of taking the ``thermal'' approach to considering quantum gravity. After all, thermodynamics is intrinsically self-limited, doesn't see microstates, and perhaps necessarily obfusicates higher order and nonperturbative corrections. However, my goal is understanding how the global structure of gravity emerges in relation to a presumably underlying, but presently unknown theory. In this sense beginning with gravity and working backwards is an ideal starting point; it has a bit of experimentally confirmed global structure to work with.

Among the other common approaches to quantum gravity there is a similar trade-off between successes and handicaps. The generally considered safest starting point is that of quantum fields in curved space \cite{preskill}. It offers a robust bulk description and has shown us the Unruh temperature of accelerated observers in flat spacetime \cite{unruh} \be \label{UH} T_{\text{Unruh-Hawking}} =\frac{\hbar \k}{2\pi} \ee and the Hawking effect \cite{hawking}. However, it is currently nonrenormalizable and due to conceptual clashes between gravity and quantum field theory, often coming down to the equivalence principle, is perhaps not the correct ultimate framework. This is of course still a discussion under development, as the \textit{asymptotic safety} program \cite{reuter}, etc, attest.

AdS/CFT \textit{defines} gravity in anti-de Sitter space in terms of a conformal field theory on an asymptotic boundary \cite{maldacena}. Because one begins with a quantum structure (albeit partially limited, as due to conformal invariance there is no Compton wavelength or mass scale) there is unitary evolution on the boundary, complemented by a partial but evolving interpretation of boundary operators in terms of bulk dynamics. One has the general \textit{GKP-Witten} postulate \cite{gkp} \cite{witten} between partition functions \be \label{GKPW} Z_{AdS_5} \equiv Z_{CFT} \simeq e^{-S_E} \ee which is complemented by a number of fascinating avenues like the long wavelength \textit{fluid/gravity} correspondence \cite{strominger1}, \textit{firewall} paradox \cite{amps1}, and more recent exploration of the \textit{ER=EPR} hypothesis \cite{maldacenasusskind1}.

In this context the Bekenstein and Wheeler \textit{Gedanken} experiments \cite{bekenstein} \cite{bekenstein2nd} \cite{bekenstein1} have a unique allure. They are consistent with the above approaches, but conceptually far more general. Their more recent implications include the holographic principle \cite{thooft} \cite{susskind1} \cite{bousso} and entropic gravity \cite{erik}. At heart they are encapsulated by the idea \be \textit{Strong Coupling} \rightarrow \textit{Gravity} \ee which is usually used in one manner or another to assign properties of a quantum system to gravity, or the holographic scaling of black holes to gauge theories.

\subsection{Black Hole \textit{Gedanken} Experiments}
The \textit{No Hair} theorem, its first versions by Israel \cite{israel1}, is essentially the progenitor of the link between thermodynamics and gravity. In thermodynamics one assigns state variables such as temperature and pressure to a system, and these decouple the effective description from the evolution of individual microstates. In the same manner the no hair uniqueness theorem characterizes black holes by state variables; their mass, angular momentum and charge, and obfuscates the behavior, or even existence, of excitations related to polarization, baryon number, quantum states, etc. 

However, all is not lost. In classical general relativity one has the Hawking \textit{area theorem} \cite{bch}, which shows the surface area of a black hole can never decrease. When multiple black holes coalesce, the combined area is greater or equal to the total initial surface area. Since the surface area is a non-decreasing quantity obeying a triangle inequality, it can be reinterpreted as an entropy, with the area theorem as the second law of thermodynamics. By analogy with the first law of thermodynamics, the surface area plays the role of entropy, and the surface gravity that of temperature. This prevents the wholesale disspearance of entropy in the vicinity of black holes. Black holes are assigned an entropy proportional to their area, an independent contribution to total entropy giving the congolomerate \textit{Generalized Second Law} of thermodynamics \cite{bekenstein2nd} \be \label{GSL} S_\text{universe} = \frac{A}{4G \hbar} + S_\text{bulk} .\ee In order to prevent more subtle violations of the generalized second law one must also posit constraints, as a result of black hole mechanics, on the entropy content of regions described by non-gravitational laws of physics. In order to see how this is realized, consider slowly lowering a particle of some mass and entropy into a black hole. As this increases the entropy of the black hole, in order to preserve the generalized second law the entropy of a region of space must have an upper bound given by the \textit{Bekenstein bound} \cite{bekenstein1} on information content\footnote{The covariant version of this bound is known as the \textit{Bousso bound} \cite{bousso}.} \be S_\text{region} \leq \frac{2\pi R E}{\hbar c} .\ee 

\subsection{\textit{Entropic} Gravity}
One of the interesting characteristics of Bekenstein's thought experiments, and the Bekenstein bound, is that at a fundamental level they apply quantum aspects to gravitational phenomena and gravitational constraints to non-relativistic and quantum systems. In the modern generalization of this, \textit{entropic gravity} \cite{erik}, the same type of logic is applied again, only this time in the flat space or Rindler approximation. The strength of this construction is that one can begin with the Bekenstein bound and reinterpret this, rather than as a relativistic bound on information content, but as an equivalently new definition of the quantum content of matter in terms of the entropy it can induce on some horizon. In the near horizon or infinite mass limit of a black hole the Schwarzschild metric becomes that of Rindler space, equivalent to an accelerated observer in flat space. This indicates that one must take seriously that, since Rindler space can be transformed to Minkowski space by a simple change of coordinates, we should also expect an entropic relation between a particle in flat space and some associated mathematical ``screen''. \be \textit{Horizon} \rightarrow \textit{Rindler} \rightarrow \textit{Minkowski} \ee This is essentially a radical reinterpretation of classical dynamics. The equivalence principle implies both gravitational \textit{and} inertial mass must have an entropic origin, if one does.

One may express the Rindler space limit of the Bekenstein bound in the curious form \be \label{EG} \D S = 2\pi k_B \frac{mc}{\hbar} \D x .\ee This is essentially the principle statement of entropic gravity; positing a complementary relation between entropy on a holographic screen, which can be in flat space, and an arbitrary but localized particle. The Compton wavelength and Boltzmann constant are added for dimensional reasons. The way in which force arises in entropic gravity, from the non-relativistic perspective, can be viewed in analogy with condensed matter. Consider a gas or fluid of molecules in a box that has a small opening, and a string placed partly inside and partly outside the box. If the string is initially straight, the stochastic motion of molecules will cause it to take a more entropically favorable arrangement, twisted and crumpled. This creates an entropic, effective force which is proportional to the entropy on the part of the string outside the box. \be \label{EF} F \D x = T \D S \ee Such a force is conservative, so long as the temperature in the box remains effectively constant. In the above example if one takes the temperature of the system to be the Unruh-Hawking temperature \ref{UH} (with the speed of light and Boltzmann constant reinserted), then \ref{EF} reduces to Newton's second law. 

This line of argument can be taken somewhat further to obtain Newton's law of gravity. The key is to assume an equipartition of energy relation of the form \be \label{EQ} E = \frac{N_{\text{bits}} k_B T}{2} . \ee From the generalized second law \ref{GSL}, the maximal entropy per surface area is given by the Bekenstein-Hawking entropy, and using this we can say that $ N_{\text{bits}}= A c^3 /G \hbar$. Substituting the Unruh temperature, \ref{EG} and \ref{EF} then gives Newton's law of gravity \be F = \frac{G m M}{r^2} .\ee

How seriously should this be taken? It depends on the weight one is willing to give to each of the individual assumptions. The Bekenstein bound and its Rindler limit are essentially a direct consequence of general relativity, and therefore a fairly solid starting point. More difficult are \ref{EF} and \ref{EQ}. One may view \ref{EF} as depending only on the entropy and Unruh temperature, and from this the force and radial response are defined as consequences. So this is again a result of the Bekenstein bound. It is then only \ref{EQ} that remains as a limiting form for equipartition, which is consistent with the limit of non-relativistic velocities. In this sense it is perfectly natural, although clearly not an ultimate building block.  

\subsection{Thermal Equilibrium --> Black Holes in AdS}
\begin{table}[tbp]
\centering
\begin{tabular}{|c|c|c|}
\hline
 & Black Hole Mechanics & Thermodynamics  \\
\hline 
Zeroth Law & $\k$ & $T$ \\
\hline
First Law & $\d M = \frac{\k }{ 8\pi G} \d A   + \O \d J$ & $\d E = T dS + \O \d J$  \\
\hline
Second Law&  $\d A \geq 0$ & $\d S \geq 0$ \\
\hline
\end{tabular}
\caption{\label{tab:i} The mapping between the laws of black hole mechanics and thermodynamics. As a result of the zeroth law, equilibrium occurs for regions of constant surface gravity, such as on the event horizon of a black hole. In the first law we identify the internal energy with $M$, temperature $T_{\text{Unruh-Hawking}}=\hbar \k / 2 \pi$, and the Bekenstein-Hawking black hole entropy $S_{\text{BH}} = A / 4 G \hbar$. The second law is a restatement of the Hawking area theorem, again with area identified as the Bekenstein-Hawking entropy.}
\end{table}
From this point I will return to general relativity. Since my goal is to apply the non-equilibrium fluctuation theorem, which is concerned with perturbations that are not necessarily at thermal equilibrium, but are between equilibrium states, to the laws of black hole mechanics, it will be helpful to have as a starting point a gravitational system in thermal equilibrium. Now the surface of a black hole has constant surface gravity, and by the laws of black hole mechanics is then in thermal equilibrium. In order to extend this to global equilibrium we must overcome the difficulty of the black hole having a negative specific heat, i.e. \be E_{\text{BH}} = \frac{1}{8\pi G T} ~~ \text{and} ~~ S_{\text{BH}}=\frac{1}{16 \pi G T}.\ee There are two main ways to do this. Consider first placing the black hole in a ``box'' with radiation distributed by the Stefan-Boltzmann law  \be E_{\text{Radiation}} = aVT^4 ~~ \text{and} ~~ S_{\text{Radiation}}=\frac{4}{3}aVT^3 .\ee This system then has a phase diagram with local and global equilibria. 

A slightly more natural way of acheiving the same aim is to replace the box with Anti-de Sitter space \cite{hawkingpage}. Here the black hole now has positive specific heat and exhibits stable equilibria for temperatures greater than \be T \geq \frac{1}{\pi} \( - \frac{\L}{3} \)^{\frac{1}{2} } . \ee These specific details will not be particularly important for applying the fluctuation theorem; the objective is to ensure its assumptions are satisfied. 

Anti-de Sitter space is further interesting in that it exhibits a form of confinement, realized through the infinite redshift of signals as they approach the AdS radius. At this level, with information concerning a mass scale or Compton wavelength redshifted away, one can construct a unitary conformal field theory on the boundary. One can also consider this radial boundary surface as a holographic screen, according to the entropic gravity prescription, however it is not entirely clear how to connect the unitary evolution of surface states to holographic screens at other radii. In what follows I show one can consider the evolution of the horizon microstates as a consequence of the Bekenstein bound or entropic gravity. I will not however construct evolution between arbitrary screens.

\section{Hawking Radiation}
Here follows a terse overview of Hawking radiation; the energy flux emitted by a black hole in, for instance, a scalar field background \cite{hawking}. In general, in order to see particle creation in a curved manifold, one can compare the Bogoliubov cofficients of field modes in the vacuum states that are seen by different observers. Since different observers will disagree about which states are vacuum states, they will also not have a shared difinition of particle states, as characterized by field excitations with postive definite norms. 

In the Schwarzschild metric one can see this effect by comparing the field modes of a scalar field in Kruskal coordinates, as seen by freely falling observers as they pass the horizon, with those seen by a constantly accelerating observer at fixed radial distance in Schwarzschild coordinates. This gives the result \be {\left| \frac{\b}{\a} \right|}^2 = \exp{\( - \frac{\o}{T_\text{UH}} \)} , \ee and requiring the normalization of field modes $<f_\text{out}, f_\text{out}>=|\a|^2 - |\b|^2=1$, we have a Bose-Einstein distribution for particle creation and annihilation \be N_\o = |\b|^2 = \frac{\G(\o)}{\exp{\( \frac{\o}{T_\text{UH}} \)} -1} . \ee Here $\G(\o)$ is a greybody factor which determines deviations of this from a blackbody spectrum. These occur due to the additional physical parameters in the problem, such as the change in mass of the black hole as it evaporates.

My initial goal in these notes is to reexpress the derivation of Hawking radiation as a non-equilbrium thermodynamics process. One would hope to eventually use such a new formulation to extract general principles on the evolution of microstates with respect to differing observers in gravity, etc. 

\subsection{Tunneling Across Horizon}
There is an interesting and more recent derivation of Hawking radiation \cite{pw} \cite{hemmingkvakkuri}, where it is explictly viewed as a semiclassical tunneling process, that I would like to highlight because of its similarity with the non-equilibrium thermodynamics viewpoint.

The idea is to reconsider the original ``heuristic'' picture of Hawking radiation as a tunneling process where pairs of particles are spontaneously created and one either tunnels out of the horizon or falls into it. This is done in the context of a Painleve line element of the form \be ds^2 = - F(M,r) dt^2 + 2 \sqrt{\frac{2M}{r}}dt dr + r^2 d\O^2  , \ee  where $F(M,r) = \( 1- \frac{2M}{r} \), $ which is regular across the horizon. 

When a quanta of radiation with energy $\o$ is emitted by the black hole, its geometry is altered through the loss of mass. If we demand conservation of energy, then the total ADM mass of the spacetime is fixed as $M$, while in the emission process geodesics near the black hole \be \dot{r} \equiv \frac{dr}{dt} = \pm 1 - \sqrt{\frac{2M}{r}} \ee change, along with the black hole metric, as $M \rightarrow M - \o$. 

The action is \be \mathcal{S} = \int^{r_\text{out}}_{r_\text{in}}  p_r dr \ee and its imaginary contribution gives the tunneling probability. By changing variables using $\dot{r}=dH/dp_r |_r$ one can compute this as an integral over the energy change \be \text{Im}(\mathcal{S}) =  \text{Im} \int^\o_0 \int^{r_\text{out}}_{r_\text{in}}  \frac{dr}{F(M,\o',r)} (-d\o') = 4 \pi \o \( M - \frac{\o}{2}  \) .\ee 

Since there are two independent processes, pair production just inside the horizon with tunneling outward, and pair production outside the horizon with an antiparticle tunneling (or falling) inward, the combined tunneling probability is approximately \be \label{THR} \G(\mathcal{S}) = \exp\( - 8\pi \o \( M - \frac{\o}{2}  \)\) = \exp\( \D S_{\text{BH}}  \) . \ee To first order this gives the classic Hawking result. 

It is interesting that the result of this semiclassical analysis is the dependence of emission probability on a purely thermal quantity, the change in black hole entropy. Also reenforcing the thermal viewpoint is the membrane paradigm-centric derivation of Hawking radiation by Parikh \cite{parikh}. As the ``stretched horizon'' decreases in size due to mass loss, its action changes to reflect this. The imaginary contribution to the membrane action is proportional to emission probability, giving the same result. 

\section{Fluctuation Theorem}
The fluctuation theorem \cite{{jarzynski}} \cite{crooks} \cite{jarzynski2} \cite{tasaki} \cite{talknerhanggi} is a general result in non-equilibrium thermodynamics which essentially depends only on the statistical reversibility of the underlying system. In the case of gravity we know that if there is an underlying system, then as a result of the form of the Bekenstein-Hawking entropy and Bousso bound, its dynamics must be time reversal invariant. We do not however know that there must actually be such a system obeying the laws of quantum mechanics; this is an assumption. The fluctuation theorem comes in two flavors, classical and quantum, appearing in essentially the same general form. However, beginning with the classical version; 

The general idea is to consider transitions between thermal equilibrium states and compare the probability density of entropy production along one path to the probability density along its time-reversed path \be \label{FT} \frac{p_{t_f, t_i}(\WW)} {p_{t_i, t_f}(-\WW) } = e^{-\b \( \D F - W \)} . \ee The usefulness of the fluctuation theorem is that the left hand side of this equation depends only on the relationship between individual microstates of the system, relations that are valid without reference to equilibrium, while the right side depends only on state variables constructed from the partition function.

A simple way to see the fluctuation theorem is through starting with the free energy in the canonical ensemble. The free energy is \be \D F = -T \D S + W ,\ee and can be rearranged to read \be e^{-\b \(\D F - W \)\ } = \frac{\O_f }{\O_i} ,\ee where $ \O_n$ is the number of states in the n$^{th}$ microstate. To link this with the fluctuation theorem, consider that if a priori we have a system with some number of A and B states, and an independent probability of transitioning between states, then if we run the system long enough we will end up with macroscopic occupation numbers for the A and B states that reflect the microscopic transition probabilities. That is, \be \frac{\O_f}{\O_i} =\frac{p_{t_f, t_i}(\WW)} {p_{t_i, t_f}(-\WW) }  . \ee In order to show this directly one can construct the free energy by averaging over the internal energy of the system, i.e. \be \left< e^{- \(  \frac{E_f}{T_f}  - \frac{E_i}{T_i} + \S_{n=1}^N \frac{\D E_n}{T_n}  \)}  \right> = \frac{Z_f}{Z_i} = \frac{\O_f}{\O_i} . \ee That this is true is a restatement of the principle of detailed balance, and it's just a coincidence that we're looking at the final and initial states; any could be used as it holds always.  So the fluctuation theorem shows that if one knows the partition function over a macroscopic equilibrium then the microscopic transition probabilities can be found.

\section{Hawking Radiation from Fluctuation Theorem}
\label{hrft}
Observing Hawking radiation from the fluctuation theorem is essentially just a straightforward application of the fluctuation theorem to the laws of black hole thermodynamics. Recall the first law of black hole thermodynamics \be \d E = T dS + \d W .\ee Here $T = \hbar \k / 2 \pi $ is the Unruh-Hawking temperature, $dS = \d A / 4 G \hbar $ the black hole entropy, and work $\d W = \O_\text{BH} \d J$. From this law one can again reconstruct the fluctuation theorem. First, defining the free energy \begin{align} \d F &= - T dS + W \\ -\b \(\d F - W  \)&= \log\(\frac{\O_f}{\O_i} \) .  \end{align} Here identifying $S = \log(\O) = p(\WW )$ is curious. What are $\O_i$? This is unfortunately presently unknown, and positing an interpretation of the thermal entropy as the log of some collection of microstates is an explicit, though probably fairly uncontroversial extra step beyond classical gravity. 

The probability of an entropy production $p_{t_f,t_i}(\WW)$ along a path $\WW$ is then as \ref{FT} \be \label{FTHR} \G(\D S_\text{BH}) = e^{-\b \( \d F - W  \) }= e^{\D S_\text{BH}} = \frac{p_{t_f, t_i}(\WW) }{p_{t_i, t_f}(-\WW)} .\ee This tells us that as the black hole metric changes, the balance of probability distributions of the entropy production along a given path, vs that of the time-reversed process, is determined by the thermal entropy change. In other words, for some entropy change between equilibrium states, where the surface area decreases (but the generalized second law is still preserved), the black hole is more likely to emit a particle than absorb one. 

This can be identified with the tunneling probability. In the context of applying the fluctuation theorem to laboratory systems the appearance of radiation through the tunneling of paths is fairly standard and, for some systems, has been observed experimentally \cite{kung}. This expectation also has the same form as \ref{THR}, which is an extra piece of evidence favoring its interpretation as black hole radiation.

\subsection{Horizon Microstates}
We do get an extra piece of information by using the fluctuation theorem, particularly if we assume 
its validity on an underlying statistical system. There are certain effects that survive in general in the thermal description, and one example of this is the time invariance of the Bousso bound \cite{bousso}. 

Returning again to the fluctuation theorem, we apply it to the laws of black hole mechanics in such situations where global thermal equilibrium can be defined, with the resulting form \ref{FTHR}.
Recall that with respect to the canonical ensemble, the free energy may be written in terms of the partition function as \be F_i = T_i \ln Z_i . \ee This is given by the Gibbons-Hawking partition function \cite{gibbonshawking} in situations where it can be well defined, like for AdS-Schwarzschild spacetimes. We could also perhaps say that $Z_{\text{AdS}} = Z_{\text{CFT}} $ by using the GKP-Witten relation \ref{GKPW} , thus gaining another way of calculating the above. However, more generally, the useful part of this partition function for calculating Hawking radiation is just the black hole entropy. 

Now, since any energy loss from the black hole causes it to decrease in area and thus entropy, \be T_{\text{BH}} \D S_{\text{BH}} = -Q_{\text{HR}} , \ee and this is carried off by the Hawking radiation, we can relate their free energies as \be \D F_{\text{BH}} = - \D F_{\text{HR}} ,\ee which can be used to write the fluctuation theorem only in terms of the Hawking radiation side. We also know that the probability of the black hole emitting a particle, and thus again producing negative entropy, must be equal to the probability of the Hawking radiation producing a corresponding positive amount of entropy. If the surface area increases, then this is the reverse process for Hawking radiation. That is, \begin{align} P_{i}(-\WW_{\text{BH}})  &= P_{f}(\WW_{\text{HR}})  \noi P_{f}(\WW_{\text{BH}}) &=P_{i}(-\WW_{\text{HR}}). \end{align} Thus, the balance of Hawking radiation is \be \frac{P_{i}(-\WW_{\text{HR}})}{P_{f}(\WW_{\text{HR}})}  = \exp(\D S_{\text{BH}})  . \ee

From entropic gravity we have the independent relation \ref{EG} \be \D S = 2 \pi k_B \frac{mc}{\hbar} \D x .\ee This is the Rindler limit of the Bekenstein bound; it is a restriction on the entropy content of matter in the vicinity of the horizon, which can be reinterpreted as a quantum statement of holography. It is perhaps not completely prescriptive, as it does not specify the global evolution, etc. Rather, it can be taken as a partial statement which should be mutually consistent on both the gravity and quantum sides. 

The Hawking radiation spectrum \ref{FTHR}, \ref{THR} is isotropic and emits along the entire angular distirbution of the black hole. If we take the Rindler limit of this expression for the purpose of comparing with \ref{EG}, then this corresponds to a small solid angle and small proper distance from the horizon approximation. The first, due to isotropy, is just a leading numerical factor on the number of states per area. The radial dependence should not affect the number of states (ignoring more complicated dynamic effects such as pair creation of massive particles which may have insufficient momentum to escape), since the quanta of Hawking radiation trace from the horizon to asymptotic infinity. 

In the near horizon limit, combining the entropic gravity prescription with the fluctuation theorem's statement on Hawking radiation is \be e^{2 \pi k_B \frac{mc}{\hbar} \D x} \propto \frac{p_i(-\WW_{\text{HR}})}{p_f(\WW_{\text{HR}})} = \frac{\O_i}{\O_f} .\ee This tells us that the number of Hawking radiation microstates evolves non-perturbatively as \be \O_f = e^{-2 \pi k_B \frac{mc}{\hbar} \D x} \O_i . \ee

\section{Discussion}
In this paper the general idea has been to extrapolate from thermal or entropic properties of black holes, through the assumption of an underlying statistical system that is consistent with one leading to the fluctuation theorem, in order to make a few initial statements as to the expected behavior of such underlying states. The principle effect is that the fluctuation theorem predicts the appearance of tunneling phenomena, and its form coincides with that independently found to be expected for Hawking radiation semiclassically. This lends the interesting perspective that applying nonequilibrium thermodynamics to gravity gives another method of naturally seeing semiclassical corrections. This can perhaps also be extended a bit further by instead using the GKP-Witten relation.

An interesting perspective is that since versions of the fluctuation theorem have been shown to hold more generally, in the absence of initial and final equilibrium states, it may be possible to use the fluctuation theorem in order to define a partition function for gravity in situations where the existence of a canonical ensemble is lacking.



\begin{thebibliography}{99}
\bibitem{israel1}
W. Israel,
\emph{Event Horizons in Static Vacuum Space-Times,}
\href{http://dx.doi.org/10.1103/PhysRev.164.1776}{Phys. Rev. 164 (5): 1776-1779}, (1967).

\bibitem{bekenstein}
J. D. Bekenstein, 
\emph{Black holes and entropy,}
\href{http://journals.aps.org/prd/abstract/10.1103/PhysRevD.7.2333}{Phys.\ Rev.\  D {\bf 7}, 2333,}, (1973).

\bibitem{bch}
J. M. Bardeen, B. Carter, S. Hawking,
\emph{Four laws of Black Hole Mechanics,}
Comm. Math. Phys. Volume 31, Number 2, 161-170, 
\href{http://projecteuclid.org/euclid.cmp/1103858973}{ProjectEuclid.org/1103858973}, (1973).

\bibitem{bekenstein2nd}
J. D. Bekenstein,
\emph{Generalized second law of thermodynamics in black-hole physics,}
\href{http://journals.aps.org/prd/abstract/10.1103/PhysRevD.9.3292}{PhysRevD.9.3292}, (1974).

\bibitem{hawking}
S. W. Hawking,
\emph{Particle Creation By Black Holes,}
\href{http://projecteuclid.org/euclid.cmp/1103899181}{Commun Math. Phys. 43, 199-220}, (1975).

\bibitem{unruh}
W. G. Unruh,
\emph{Notes on Black Hole Evaporation,}
\href{http://journals.aps.org/prd/abstract/10.1103/PhysRevD.14.870}{PhysRevD.14.870}, (1976).

\bibitem{gibbonshawking}
G. Gibbons, S. W. Hawking,
\emph{Action integrals and partition functions in quantum gravity,}
\href{http://journals.aps.org/prd/abstract/10.1103/PhysRevD.15.2752}{PhysRevD.15.2752}, (1977).

\bibitem{bekenstein1}
J. D. Bekenstein,
\emph{Universal upper bound on the entropy-to-energy ratio for bounded systems,}
\href{http://prd.aps.org/abstract/PRD/v23/i2/p287_1}{Phys. Rev. D, 23, No. 2, 287-298}, (1981).

\bibitem{hawkingpage}
S. W. Hawking, D. N. Page,
\emph{Thermodynamics of black holes in anti-de Sitter space,}
Comm. Math. Phys. 87 (1982), no. 4, 577-588
\href{http://projecteuclid.org/euclid.cmp/1103922135}{ProjecteEuclid.org/1103922135}, (1982).

\bibitem{pricethorne}
R. H. Price and K. S. Thorne,
\emph{Membrane Viewpoint on Black Holes: Properties and Evolution of the Stretched Horizon,}
\href{http://link.aps.org/doi/10.1103/PhysRevD.33.915}{Physical Review D, 33, 915-941}, (1986). 

\bibitem{preskill}
J. Preskill,
\emph{Quantum Field Theory in Curved Spacetime,}
Physics 236c, Quantum Field Theory in Curved Spacetime notes, (1990).

\bibitem{thooft}
G.~'t Hooft,
\emph{Dimensional reduction in quantum gravity,}
\href{http://arxiv.org/abs/gr-qc/9310026}{arXiv:9310026}, (1993).

\bibitem{susskind1}
L. Susskind, 
\emph{The World as a hologram,}
  J.\ Math.\ Phys.\  {\bf 36}, 6377
\href{http://arxiv.org/abs/hep-th/9409089}{arXiv:9409089}, (1995).

\bibitem{jarzynski}
C. Jarzynski,
\emph{A nonequilibrium equality for free energy differences,}
Phys Rev Lett 78, 2690 (1997),
\href{http://arxiv.org/abs/cond-mat/9610209}{arXiv:cond-mat/9610209}, (1996).

\bibitem{maldacena}
J. M. Maldacena,
\emph{The large N limit of superconformal field theories and supergravity,}
Adv. Theor. Math. Phys. 2, 231, (1998)
Intl. J. Theor. Phys. 38, 1113, (1999).
\href{https://arxiv.org/abs/hep-th/9711200}{arXiv:hep-th/9711200}, (1997).

\bibitem{gkp}
 S. S. Gubser, I. R. Klebanov and A. M. Polyakov, 
\emph{Gauge theory correlators from noncritical string theory,}
Phys. Lett. B428 (1998) 105
\href{http://arxiv.org/abs/hep-th/9802109}{arXiv:hep-th/9802109}, (1998).

\bibitem{witten}
 E. Witten,
\emph{Anti-de Sitter space and holography,} Adv. Theor. Math. Phys. 2 (1998) 253 
\href{http://arxiv.org/abs/hep-th/9802150.pdf}{arXiv:hep-th/9802150}, (1998).

\bibitem{parikh}
M. Parikh,
\emph{Membrane Horizons: The Black Hole's New Clothes,}
Ph.D. Thesis. Princeton University, October 1998
\href{http://arxiv.org/abs/hep-th/9907002}{arXiv:hep-th/9907002}, (1998).

\bibitem{crooks}
G. E. Crooks,
\emph{The Entropy Production Fluctuation Theorem and the Nonequilibrium Work Relation for Free Energy Differences,}
Phys. Rev. E 60, 2721 (1999),
\href{http://arxiv.org/abs/cond-mat/9901352v4}{arXiv:cond-mat/9901352}, (1999).

\bibitem{pw}
M. Parikh, F. Wilczek,
\emph{Hawking Radiation As Tunneling,}
Phys.Rev.Lett.85:5042-5045,2000,
\href{http://arxiv.org/abs/hep-th/9907001v3}{arXiv:9907001v3}, (1999).

\bibitem{jarzynski2}
C. Jarzynski,
\emph{Hamiltonian derivation of a detailed fluctuation theorem,}
LAUR-99-2903,
\href{http://arxiv.org/abs/cond-mat/9908286}{arXiv:cond-mat/9908286}, (1999).

\bibitem{tasaki}
H. Tasaki,
\emph{Jarzynski Relations for Quantum Systems and Some Applications,}
\href{http://arxiv.org/abs/cond-mat/0009244}{arXiv:cond-mat/0009244}, (2000).

\bibitem{hemmingkvakkuri}
S. Hemming, E. Keski-Vakkuri,
\emph{Hawking Radiation from AdS Black Holes,}
Phys.Rev.D64:044006,2001,
\href{http://arxiv.org/abs/gr-qc/0005115v2}{arXiv:gr-qc/0005115}, (2000).

\bibitem{bousso}
R. Bousso,
\emph{The Holographic Principle,}
Rev.Mod.Phys.74:825-874,2002,
\href{http://arxiv.org/abs/hep-th/0203101v2}{arXiv:hep-th/0203101v2}, (2002).

\bibitem{talknerhanggi}
P. Talkner, P. Hanggi,
\emph{The Tasaki-Crooks quantum fluctuation theorem,}
J. Phys. A 40, F569-F571,
\href{http://arxiv.org/abs/0705.1252v1}{arXiv:0705.1252}, (2007).

\bibitem{erik} 
E. Verlinde,
\emph{On the Origin of Gravity and the Laws of Newton,}
Journal of High Energy Physics: 1104:029,
\href{http://arxiv.org/abs/1001.0785}{arXiv:1001.0785}, (2011).

\bibitem{strominger1}
I. Bredberg, C. Keeler, V. Lysov, A. Strominger,
\emph{From Navier-Stokes To Einstein,}
\href{http://arxiv.org/abs/1101.2451}{arXiv:1101.2451}, (2011).

\bibitem{reuter}
M. Reuter, F. Saueressig,
\emph{Quantum Einstein Gravity,}
New J.Phys. 14 (2012) 055022,
\href{http://arxiv.org/abs/1202.2274.pdf}{arXiv:1202.2274}, (2012).

\bibitem{amps1}
A. Almheiri, D. Marolf, J. Polchinski, J. Sully,
JHEP, 10.1007/JHEP02(2013)062
\emph{Black Holes: Complementarity or Firewalls?,}
\href{https://arxiv.org/abs/1207.3123}{arXiv:1207.3123}, (2012).

\bibitem{maldacenasusskind1}
J. Maldacena, L. Susskind,
\emph{Cool horizons for entangled black holes,}
Fortsch.Phys. 61 (2013) 781-811,
\href{https://arxiv.org/abs/1306.0533}{arXiv:1306.0533}, (2013).

\bibitem{kung}
B. Kung, et al.
\emph{Test of the fluctuation theorem for single-electron transport},
\href{http://dx.doi.org/10.1063/1.4795540}{J. Appl. Phys. 113, 136507}, (2013). 

\bibitem{natsuume}
M. Natsuume, 
\emph{AdS/CFT Duality User Guide},
Springer,
\href{http://arxiv.org/abs/1409.3575}{arXiv:1409.3575v3}, (2015).

\bibitem{harlow1}
D. Harlow,
\emph{Jerusalem Lectures on Black Holes and Quantum Information,}
Rev. Mod. Phys. 88, 15002 (2016),
\href{http://arxiv.org/abs/1409.1231}{arXiv:1409.1231}, (2016).





\end{thebibliography}
\end{document}